\documentclass[traditabstract]{aa}
\usepackage{graphicx}
\usepackage{subcaption}
\usepackage{hyperref}
\usepackage{txfonts}
\usepackage{amsmath}
\usepackage{color}

\begin{document}

   \title{Can a negative-mass cosmology explain dark matter and dark energy?}


   \author{H. Socas-Navarro
          \inst{1} 
          }

   \institute{Instituto de Astrof\'\i sica de Canarias,
     Avda V\'\i a L\'actea S/N, La Laguna, E-38205, Tenerife, Spain
     \and
     Departamento de Astrof\'\i sica, Universidad de La Laguna, La Laguna, E-38205, Tenerife, Spain \\
     \email{hsocas@iac.es}
   }
   
   \date{Received February 5, 2019; accepted }

 
  \abstract
  {A recent study by Farnes (2018) proposed an alternative
    cosmological model in which both dark matter and dark energy are
    replaced with a single fluid of negative mass. This paper presents
    a critical review of that model. A number of problems and
    discrepancies with observations are identified. For instance, the
    predicted shape and density of galactic dark matter halos are
    incorrect. Also, halos would need to be less massive than the
    baryonic component, otherwise they would become gravitationally
    unstable. Perhaps the most challenging problem in this theory is
    the presence of a large-scale version of the ``runaway effect'',
    which would result in all galaxies moving in random directions at
    nearly the speed of light.
    Other more general issues regarding negative mass in
    general relativity are discussed, such as the possibility of
    time-travel paradoxes.  }


   \maketitle
%

\section{Introduction}
\label{sec:intro}

In a recent paper, \cite{F18} (hereafter F18) put forward an audacious
proposal to explain dark matter and dark energy in a unified scenario
with a single fluid of negative-mass particles. This fluid would be
continuously created everywhere to maintain its density at a constant level in
spite of cosmological expansion. To support his hypothesis, F18
presented a number of simple numerical simulations showing the
formation of halos around galaxies and indications of filament and
void structures in a cosmological volume. In the spirit of healthy
scientific debate, this paper presents the results of a critical look
at the F18 model and some objections derived from the analysis of
similar simulations.

The idea of matter with negative mass is not new in physics. In fact,
some speculative work dates back to the nineteenth century (see e.g.,
the introduction of \citealt{B89}). After the development of general
relativity, when physicists were struggling with the conflict between
the crunching action of a universal gravity and the prevailing notion
at the time of a static universe, negative mass was seen by some authors as a
promising solution to one of the biggest problems in cosmology. A
repulsive form of gravity emanating from negative masses could provide
the necessary force to counterbalance the universal collapse
(\citealt{E18}). In reality, this statement is deceivingly simple and
many important complications lurk behind it. To begin with, one has to
define exactly what is meant by negative mass and this is not
straightforward. In general relativity, the source of gravitation is
the energy-stress tensor, a complex set of quantities that involves
the density and flux of mass, energy, and stress. Remarkably, different
observers in different reference frames might even measure opposite
signs of some of these quantities.

It is tempting to draw a parallelism between a dual-sign gravitation
(with positive and negative masses) and the dual-charge nature of
electromagnetism. However, this would be a very na\"\i ve conception,
as the two situations are completely different. The crucial difference
is that, in the case of electromagnetism, two equal (opposite) charges
repel (attract) each other, whereas in a dual-sign gravitation the
opposite process would take place. The difference may appear subtle but
it is of paramount importance. Gravity is able to attract many
particles of the same sign together and make them coalesce into
arbitrarily large structures.

In general relativity, the solution to the simple scenario where we
have a point-like mass in an otherwise empty Minkowskian space-time is
the \cite{S16} metric. This solution has a singularity at the mass
position but the singularity is hidden from external observers by an
event horizon that surrounds it. No information may propagate from the
singularity to the outside world. The equivalent of the Schwarzschild
solution for a negative mass does not represent a symmetric
situation. It also has a singularity at the mass position but now
there is no horizon. The singularity is naked, in violation of the
weak cosmic censorship conjecture (\citealt{P69}). While the cosmic
censorship remains an unproven conjecture, it is widely recognized by
the community as a sensible hypothesis and solutions with naked
singularities are generally regarded as unsatisfactory. Of course,
point-like masses do not exist in reality but it is often the case
that one may mathematically construct a continuous mass distribution
by integration of infinitesimal points. This raises yet another
problem. If we consider a finite-size negative mass object, how are
its pieces held together? In principle, each portion of it would be
subject to a repulsive force from the rest of the object, with the
force increasing to infinity for decreasing distances. Particles with
electric charge are held together by quantum physics and the existence
of an elementary charge. However, there is no such thing for gravity,
at least in the context of general relativity. All of these are
conceptual problems which, while not necessarily ruling out the
existence of negative mass in the Universe, should at least serve as a
warning sign that it is a complex issue and certainly not a symmetric
counterpart of ordinary matter.

The concept of mass may have four different meanings in physics: the
ability of matter to produce gravity (``active'' gravitational mass),
the response of matter to gravity (``passive'' gravitational mass),
the inertia of matter (its resistance to accelerate when subject to a
force), or its energy equivalence. Classically, all of these
definitions of mass refer to the same quantity and are positive by
definition. General relativity requires by construction (by virtue
of the equivalence principle) that passive gravitational and inertial
mass must be the same. The theory does not explicitly require that they be positive, although some restrictions exist, as discussed
in Sect.~\ref{sec:others} below.

Various ideas were proposed and discussed in the decades following the
development of general relativity regarding the interpretation and
implications of negative sources of gravitation in that
theory. Critically important was the work of \cite{B57}. First
\cite{L51} and then \cite{B57} (using Newtonian and relativistic
frameworks, respectively), studied the concept of what is known as
the ``runaway motion'': two particles with opposite mass would start
accelerating in one direction. In the Newtonian framework, this
happens because the negative mass is attracted towards the positive
one (the force is in the opposite direction but the negative inertial
mass makes it accelerate towards the positive mass), whereas the
positive mass is repelled and accelerates away from the negative.  The
same effect occurs in general relativity.

Following these developments, interest in the physics of negative
masses declined, at least as a practical idea (exploratory theoretical
works were still being pursued), for two main reasons. First, the
runaway effect was considered so ``preposterous'' that it was viewed
by many authors as unphysical. Second, the discovery of the
Hubble-Lema\^\i tre law (\citealt{L27}; \citealt{H29}) and the later
confirmation of the Big Bang model with the detection of the cosmic
microwave background (\citealt{PW65}) removed what had been the
original motivation for a repulsive gravitational force, that is, a means
to sustain a steady-state universe (e.g., \citealt{BG48}).

However, the disinterest did not last long. A renewed interest in the
issue of negative mass was sparked by the recent discovery of the
accelerated cosmological expansion and its association with a
mysterious dark energy and the cosmological constant
(\citealt{RFC+98}; \citealt{PR03}).

Modern negative-mass cosmologies typically attempt to explain dark
energy as a repulsive form of gravity. \cite{H08} and \cite{PdA14}
explored an extension of general relativity with two different
metrics, with positive and negative mass distributions, respectively
(and a different speed of light). \cite{BLC12} proposed the so-called
Dirac-Milne universe, a cosmology combining positive-mass matter in a
sea of negative-mass anti-matter (see also \citealt{MGR+18}). It
is worth noting that, while anti-matter has positive mass in the
standard model of particle physics, this prediction has not yet been
tested experimentally due to the overwhelmingly larger effect of the
electromagnetic force compared to gravity at the atomic
scale. Experiments are currently underway that may soon resolve this
question (see, e.g., \citealt{ABB+19}), at least for the passive
gravitational mass. In any case, the inertial mass of anti-matter must
be positive, given that its acceleration has been measured in
electro-magnetic fields.

The negative-mass cosmology proposed by F18 is fundamentally new in
that it replaces both dark matter and dark energy with one single
ingredient, namely the negative-mass fluid. His claim is sustained on
numerical simulations of how this fluid would operate in different
physical processes, such as galaxy evolution or cosmological
collapse. The simulations presented in F18 are encouraging and show
remarkable agreement with the observations (formation of halos, flat
galactic rotation curves, cosmological structures, etc.). While we have
rather robust constraints on the mass of dark matter concentrations by
means of gravitational lensing observations, we are not able in
most situations to determine the sign of the lensing mass
(\citealt{TA13}). If the F18 ideas were proven correct, the
implications for all of physics would be formidable. A close
scrutiny of this model is therefore clearly warranted.

This paper presents a critical analysis of the F18 results,
identifying a number of problems and incompatibilities with
existing observations. Section~\ref{sec:analytical} has an analytical
derivation of the expected properties of galactic halos in the F18
scenario. This will be helpful to understand and validate some of the
results in Sect.~\ref{sec:sims}, which presents simulations and the
discrepancies with observations. Section~\ref{sec:others} emphasizes
other conceptual difficulties facing the model that are not directly
testable against observations. Finally, the overall conclusions are
summarized in Sect.~\ref{sec:conclusions}.

\section{Galactic halos}
\label{sec:analytical}

One of the remarkable results of F18 was the formation of negative
mass halos around galaxies with a density structure strikingly
similar to a Navarro-Frenk-White (NFW, \citealt{NFW96}) dark matter
profile. This paper presents new simulations very similar in nature to
those of F18 but the resulting halo densities exhibit what appears to
be an exponential stratification instead. In order to develop a better understanding of
the simulation results, or even to serve as a
sanity check, it is usually helpful to conduct an analytical study of
the system. In this section we seek the functional form of the density
profile of a spherical negative mass halo surrounding a central core
of positive masses.

Throughout this section we assume spherical symmetry, noninteracting
particles (except for gravitation), and a steady-state
equilibrium in which the system has reached a stable macroscopic
configuration. The radial density profiles for the positive mass core
and the negative mass halo are arbitrary and may overlap. For
simplicity, let us consider a maximum radius $R^C$ and $R^H$ for the
core and halo, respectively ($R^C < R^H$). We assume that the negative
masses are initialized with a constant density $\rho_i^H$ and no
initial velocity. This last assumption is probably not very realistic
but that is how the F18 simulations are initialized and those starting
conditions are mimicked in the present work, as well.

At any given time, the amount of halo particles contained in a
spherical shell of radius $r$ and thickness $dr$ is given by the
number of particles that have fallen from higher layers, balanced by
those that have fallen from the opposite side, crossed the origin, and
are moving upwards through $r$. Particles in layers below $r$ will never
reach this height. Consider another shell of radius $r'$ and thickness
$dr'$.  The probability that a particle that originated at shell $r'$
is inside shell $r$ at any given time may be calculated as the time
spent by such a particle within shell $r$ divided by the total time it
takes to fall all the way to the center:
\begin{equation}
  P(r',r) dr = \begin{cases}
    0 & r' < r \\
    d \tau / T &  r < r' < R^H
  \end{cases}
,\end{equation}
with
\begin{equation}
  d \tau={dr \over v(r',r)} \, ,
\end{equation}
and
\begin{equation}
  T=\int_0^{r'} {dr''\over v(r',r'')} \, .
\end{equation}
In the expressions above, $v(r',r)$ is the velocity that particles
originating at $r'$ have when they reach position $r$. Our initial
conditions require $v(r',r') = 0 \, \forall r'$. We can then write the following equation for the halo density profile $\rho^H(r)$:
\begin{multline}
  \label{eq:rho1}
  4\pi r^2 \rho^H(r) dr= \int_r^{R^H} \rho_i^H 4 \pi r'^2 P(r',r) dr' dr = \\
  \int_r^{R^H} \rho_i^H 4 \pi r'^2 {1 \over v(r',r)} {1 \over {\int_0^{r'}{dr''/v(r',r'')}}}  dr' dr \, .
\end{multline}

Since both $\rho^H$ and $v$ are unknown, another relationship is needed to
close the system, accounting for the action of gravity. The
gravitational potential of an arbitrary mass distribution is given by
Poisson's equation:
\begin{equation}
  \label{eq:poisson}
  \nabla^2 \varphi = 4\pi\rho \, .
\end{equation}
Given the symmetry of our problem, we may take advantage of the
Gaussian theorem\footnote{Also credited to M.~V. Ostrogradsky or
  G. Green. Sometimes also known as the divergence theorem.}
(\citealt{G1813}), or even the shell theorem (\citealt{N1833}), to
express the gravitational field as:
\begin{equation}
  g(r)=-G { m(r) \over r^2} \, ,
\end{equation}
where $g(r)$ is the field, $G$ is the universal gravitational constant,
and $m(r)$ is the total mass encircled by a sphere of radius $r$. This
expression, along with the force derived from the field, remains valid
in the Newtonian negative-mass formalism of F18, with $m(r)$
accounting for masses of both signs.

For both positive and negative masses, the acceleration $a$ is equal
to the field $g$. Replacing $dt$ with $dr/v$ in the relation $a=dv/dt$
and using $a=g$, we conclude that:
\begin{equation}
  vdv=-G{m(r) \over r^2} dr \, ,
\end{equation}
which may be integrated between two shells $r$ and $r'$ to yield:
\begin{equation}
  \label{eq:vel1}
  v^2(r',r)=-2G\int_{r'}^r{ m(r'') \over r''^2} dr'' \, .
\end{equation}

Since $m(r)=\int_0^r 4\pi r'^2\rho(r') dr'$, and $\rho$ is the addition of
the core and halo mass densities, Eqs.~(\ref{eq:rho1})
and~(\ref{eq:vel1}) constitute a complete system whose solutions may
be explored to investigate the properties of the density and velocity
profiles.

Motivated by the results presented in Sect.~\ref{sec:sims} below, we
try an exponential profile ansatz for the halo density:
\begin{equation}
  \rho^H(r)=-\rho_0^He^{-kr} \, .
\end{equation}
The total halo mass encircled below a certain radius $r$ may be
calculated after some tedious but straightforward series of
integrations by parts:
\begin{equation}
  m^H(r)=-4\pi\rho_0^H{2-(k^2r^2+2kr+2)e^{-kr} \over k^3 } \, .
\end{equation}
At any point in the halo $r>R^C$, the total (core and halo) mass
encircled inside $r$ is then $m(r)=m^H(r)+M^C$ (with $M^C$ denoting
the total core mass). Substituting in Eq.~(\ref{eq:vel1}) and
rearranging terms we obtain:
\begin{multline}
  \label{eq:vel2}
  v^2(r',r)=-2G\left [ \int_{r'}^r {M^C \over r''^2} dr'' + \right . \\
    {4\pi \rho_0^H \over k^3}\left (
    {2(1-e^{-kr'})\over r'} - 
    \left . {2(1-e^{-kr})\over r} + k(e{-kr}-e^{-kr'}) \right ) \right ] \, .
\end{multline}
The first term in Eq.~(\ref{eq:vel2}) accounts for the gravity
produced by the core and has a simple form because it is always
underneath. By virtue of the Gaussian or shell theorems, the field
outside is equivalent to that of a point mass at the center. 
The second term is due to the gravitational
field produced by the halo itself. If we take this equation and insert
it into Eq.~(\ref{eq:rho1}) we obtain the solution of the
system.

We note that $v(r',r)$ appears in denominators in Eq.~(\ref{eq:rho1}),
which means that it must be nonzero for the system to have a physical
solution. Moreover, Eq.~(\ref{eq:vel2}) gives us an expression for the
square of $v(r',r)$. Therefore, our system admits a real solution only
if the right-hand side of Eq.~(\ref{eq:vel2}) is strictly positive.  If
we can prove that this is the case, then our ansatz is correct and an
exponential density profile is a solution to our equations.

The first term, proportional to $M^C$, is trivially positive
(recall that $r'$ is greater than $r$ in the integration
  limits). The second term is more complex. With some rearrangement,
our problem is equivalent to proving that $\forall r < r'$:
\begin{equation}
  \label{eq:cond1}
   {2(1-e^{-kr})+kre^{-kr} \over r } -
   {2(1-e^{-kr'})+kr'e^{-kr'} \over r' }  > 0 \, .
\end{equation}
The above statement is equivalent to proving that the following
function is monotonically decreasing:
\begin{equation}
  f(u)={2(1-e^{-u})+ue^{-u} \over u} \, .
\end{equation}
Taking the derivative of $f(u)$, we can deduce that $f'(u)$ is zero
only if $u\to\infty$ or if:
\begin{equation}
  \label{eq:eu}
  e^u=-{u^2 \over 2}+u+1 \, .
\end{equation}
 We can prove that Eq.~(\ref{eq:eu}) has no solution for $u>0$ because
 in this domain the left-hand side $e^u$ is always greater than 1,
 whereas the right-hand side is a convex parabola with a maximum value
 of $3/4$. In other words, we have proven that the function $f'(u)$
 has no roots, which means that $f(u)$ is monotonic (for $u>0$). Now
 all that is left is to prove that $f'(u)$ is negative somewhere, which
 is easy to see by evaluating the function. For large values of $u$,
 it is obvious that $f'(u)$ is negative.

 We have proven that $f(u)$ is monotonically decreasing, which
 automatically means that the condition expressed in
 Eq.~(\ref{eq:cond1}) is true for all $r < r'$ and that the right-hand
 side of Eq.~(\ref{eq:vel2}) is positive. We have therefore proven that
 Eq.~(\ref{eq:rho1}) has a real solution and that the ansatz is
 correct: An exponential profile of the form $\rho^H(r)=-\rho_0^H
 e^{-kr}$ is a solution to our system.

\section{Numerical simulations}
\label{sec:sims}

One of the most interesting novelties of F18 is that it presents the
first negative-mass cosmological model to be implemented in actual
numerical simulations. The author generously made his simulation code
available to the community, a practice that will hopefully become
increasingly widespread among researchers as it greatly facilitates
verification and reproducibility of results. The importance of these
two elements in the scientific method cannot be overstated. The
results discussed in this section have been obtained with my own
simulation code, which follows the philosophy of F18 as closely as
possible but with some relevant differences, as explained in this
section.

In the F18 code, the gravitational force between two masses
  situated at a distance $d$ drops as the inverse of $d$, instead of
  the Newtonian form $d^2$, resulting in a stronger force at large
  distances. Both forms produce qualitatively similar features, such
  as halo formation or cosmological collapse, obviously with
  quantiative differences. Assessing the influence of this peculiar
  gravity on the F18 results is beyond the scope of this paper. We
  adopt a realistic Newtonian gravity with a $d^2$ dependence.

  Although F18 claims to use a ``leapfrog'' integration scheme, his
  algorithm seems to operate as a first-order Euler integrator. I have
  implemented both schemes and verified that there are no significant
  qualitative or quantitative differences in the results obtained. The
  runs discussed in this paper have been computed with the leapfrog
  scheme.


The code employed here has been developed from scratch with the
specific purpose of conducting these simulations. The source code,
written in Fortran90 with MPI (message-passing interface,
\citealt{mpi94}) parallelization, is publicly
available\footnote{\href{https://github.com/hsocasnavarro/nbody_sim}{https://github.com/hsocasnavarro/nbody\_sim}}. The
main difference with F18, apart from the $d^2$ dependence of the
  gravitational force, resides in the boundary conditions. The
  boundary treatment was not explicitly discussed in F18 but it may be
  inferred by analyzing the source code. The simulations presented in F18 are
initialized in a cubic box in which all particles are placed. The box
has open boundaries and the particles are free to spread away from it
throughout a much larger box. This second box has reflecting boundary
conditions. Any particle that reaches one of its sides will bounce
back by flipping the sign of the velocity vector component in the
corresponding direction. In practice however, the second box is so
much larger than the first one that very few particles ever reach
its boundary within the time span of the simulations. Thus, for all
practical purposes, we can consider the F18 simulations as defined in
the initial box with open boundary conditions.

In cosmological simulations, it is customary to consider the opposite
approach, with periodic boundary conditions. Periodic conditions are
suitable in cosmology because they ensure that the Universe does not
end abruptly outside the simulation domain. They produce simulations
that are automatically compliant with the cosmological principle, in
the sense that the box is a typical region and, in particular, there
is no center of the Universe. In contrast, a simulation with open
boundary conditions has a preferred direction and all matter tends to
fall towards (or be repelled from, in the case of repulsive gravity)
the center of the box. For instance, if we consider a homogeneous sea
of negative-mass particles, open boundary conditions will make the
initial box expand, as all particles repel each other into the empty
space around it. With periodic boundary conditions, the fluid
remains stable as every particle feels the same repulsive force
from all directions, even those close to the edge of the domain. 
  Another difference with modern large-scale simulations is the
  neglect of cosmological expansion, which is typically accounted for
  by adopting comoving coordinates and a modified potential. Cosmic
  expansion is neglected here in order to keep the simulations as
  close as possible to those of F18. 

\subsection{Halo mass}
\label{sec:lighthalos}

\begin{figure}
  \centering
  \begin{subfigure}[b]{0.45\textwidth}
  \includegraphics[width=\textwidth]{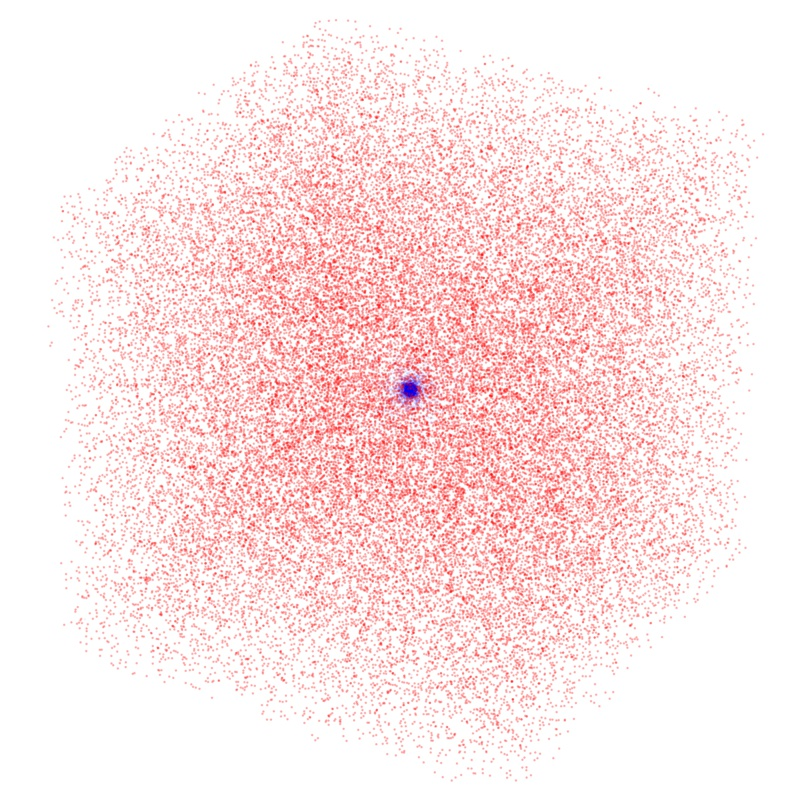}
\end{subfigure}
  \centering
\begin{subfigure}[b]{0.45\textwidth}
  \includegraphics[width=\textwidth]{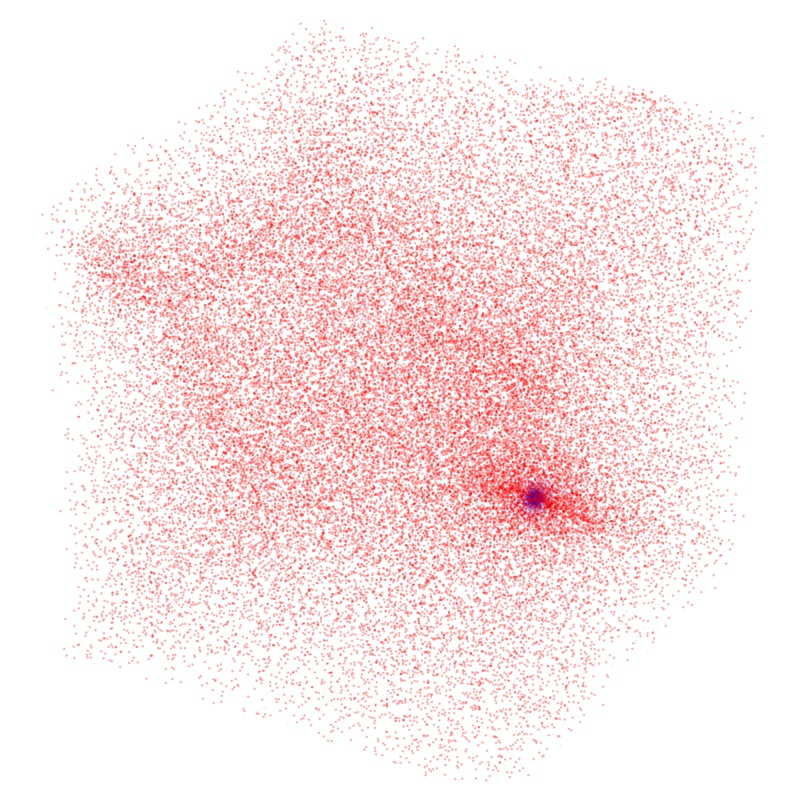}
\end{subfigure}
\caption{
    \label{fig:simhalo}
    Simulation of galactic halo formation.  Blue (red) dots
    represent the positive (negative) mass particles.  Upper image:
    Starting conditions, representing a Hernquist positive mass
    spherical galaxy embedded in a uniform distribution of negative
    masses. Lower image: Final state, exhibiting a nonspherical halo
    of enhanced negative-mass density around the galaxy. The galactic
    center of mass is initially at rest. In the final frame it is
    moving towards the left of the image. We note that the figure is
      represented in a 3D isometric perspective. Apparent particle
      densities may be misleading due to projection effects. }
\end{figure}

Let us begin by considering the galaxy halo formation, which according
to F18 plays the role of dark matter in his unified model. I employed
the same initial configuration and simulation parameters but using
periodic boundary conditions and the correct form for the newtonian
gravitational force. Figure~\ref{fig:simhalo} shows the starting and
final frames of the simulation, using the same color scheme and
isometric perspective as F18 to facilitate comparisons. The initial
configuration has a spherical distribution of positive masses (the
core) following the model of \cite{H90} embedded in a uniform sea of
 particles with negative mass.

This and other similar simulations produce halos of negative masses
around the galaxies. However, they exhibit important differences with
respect to the properties of real dark matter halos. To begin with,
the halos in these simulations are too light. The ratio of halo to
core mass in the simulations is between 0.3 and 0.8, whereas real
galaxies have dark matter halos that are typically a factor of four to five
more massive than their baryonic components. Furthermore, it is easy
to intuitively understand why negative-mass halos must be light. Halo
particles repel each other; they are held together by the attractive
force from positive masses at the center. Invoking again the shell
theorem (and assuming spherical symmetry), halo particles will only
feel a central attractive potential if the total encircled mass is
positive. If a halo had a ratio greater than one, the total mass
encircled by the outer layers would be negative and such layers would
be pushed away from the galaxy. This behavior is confirmed by
simulations in which a galaxy is initialized with a massive halo. The
halo is unstable and the outer layers are rapidly ejected from the
galaxy.

\subsection{Density profile and rotation curve}
\label{sec:density}

\begin{figure}
  \resizebox{\hsize}{!}{\includegraphics{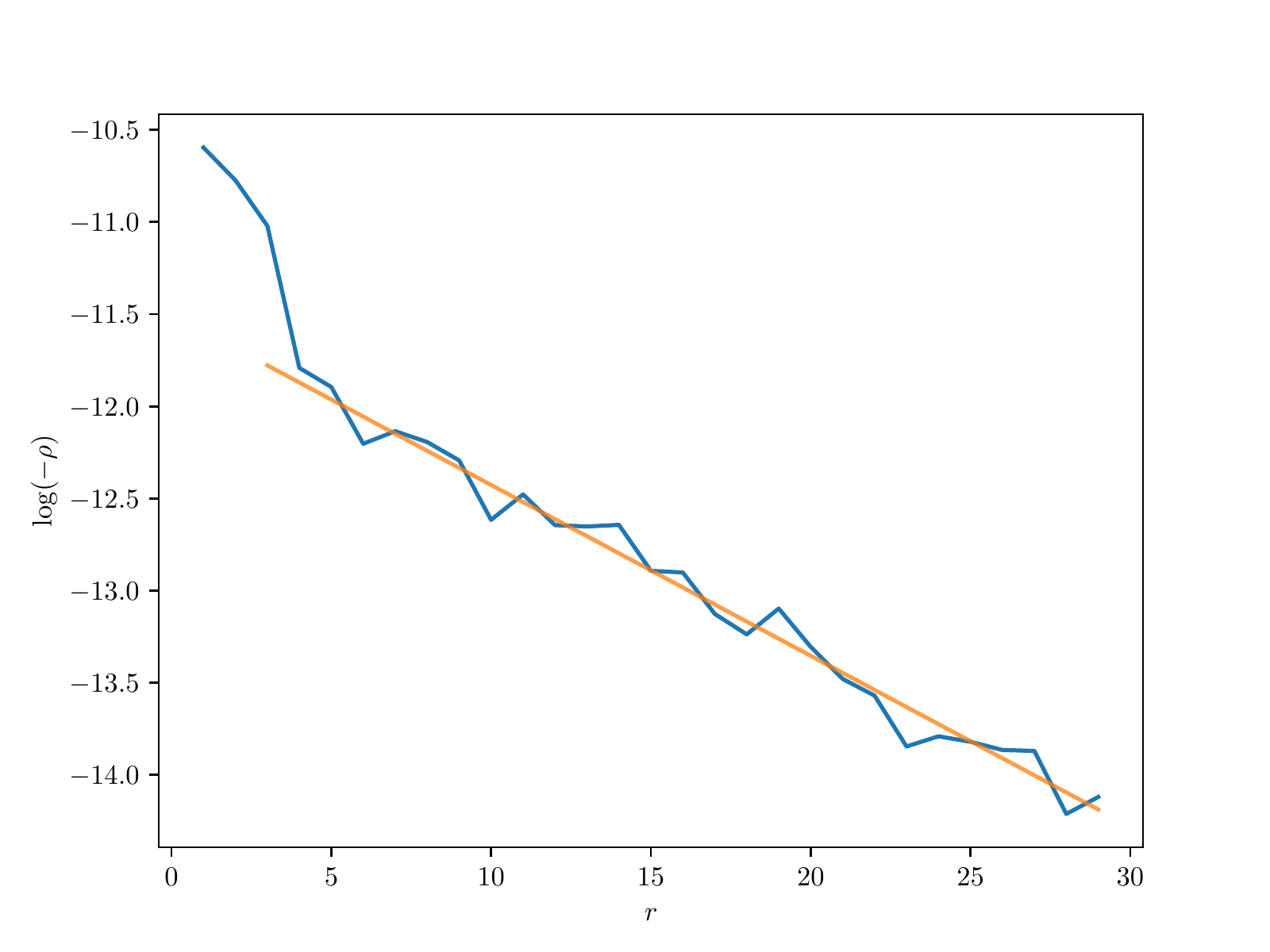}}
    \caption{ \label{fig:density} Radial density profile of the
      negative-mass halo in the galactic simulation. The
      axis of ordinates is logarithmic. The orange line shows an
      exponential density profile for reference. }
\end{figure}

The radial density profile of the negative-mass halo is shown in
Fig.~\ref{fig:density}. Outside the positive mass core, the simulation
is in good agreement with the results of Sect.~\ref{sec:analytical}
of an exponential density profile (orange line). Dark matter halos on
the other hand exhibit a more complex radial dependence (see, e.g.,
\citealt{NFW96}). 

\begin{figure}
  \resizebox{\hsize}{!}{\includegraphics{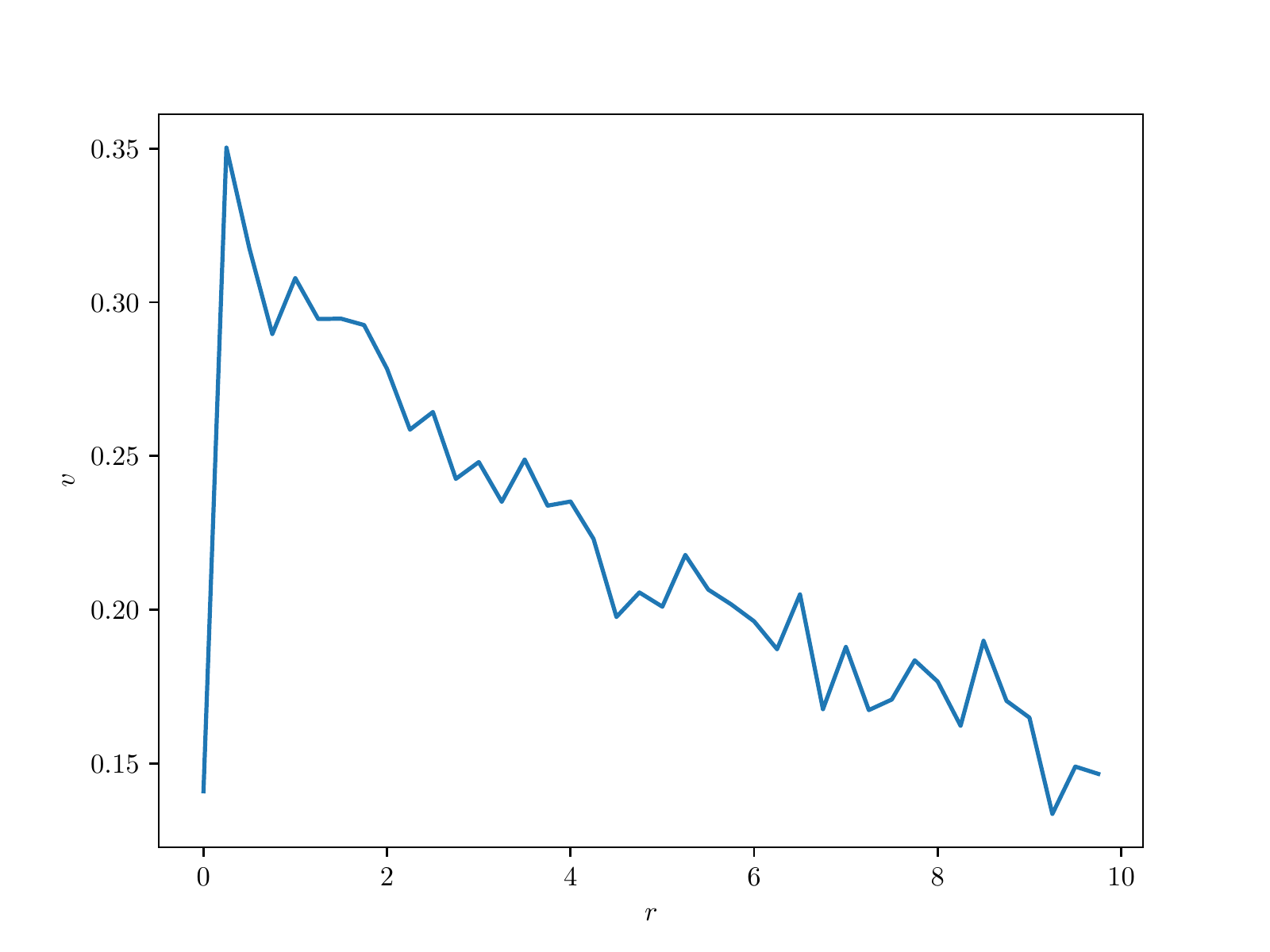}}
  \caption{ \label{fig:rotation} Average orbital velocity as a
    function of radial distance to the galactic center.  }
\end{figure}

Another discrepancy with the F18 results (and with observations) is
the galactic rotation curve. Without dark matter, the orbital velocity
of stars should decrease as we move away from the galactic
center. However, observations clearly show a flattening of this curve,
in such a way that stars move with roughly the same linear velocity
independently of their radial position. This observation is considered
as one of the first pieces of  historical evidence of dark matter
(\citealt{R70}). F18 claims that his simulation produces a flat
rotation curve and argues that this is evidence in favor of the
negative-mass model as an alternative to dark matter. However, this
simulation exhibits different stellar
dynamics. Figure~\ref{fig:rotation} shows that the orbital velocity of
stars drops with distance to the center, in disagreement with the
observed flattened rotation curves.
  
\subsection{Accelerating galaxies}
\label{sec:runaway}

Perhaps the most important problem with the negative-mass simulations
is that a galaxy, whose center of mass is initially at rest,
immediately starts accelerating in a random direction and continues to
gain speed during the entire run. This odd acceleration is present in
all of the galactic simulations performed and, again, may be
understood intuitively, in this case as a large-scale version of the
runaway effect.

Let us imagine an isolated system consisting of a positive-mass
  particle standing at the center of a spherical
  negative-mass halo. The total force exerted over the positive mass
is exactly zero because the forces from opposite directions balance
each other out. However, this equilibrium is unstable in the sense
that if the particle position is perturbed slightly, there would be
more negative particles pushing from behind and fewer pushing from the
front. Thus, there would be a net force pushing in the direction of
the motion. Mathematically, the stability condition requires that the
gravitational potential be concave at the origin, that is, a
positive second derivative. Since by definition $\vec g(\vec
r)=-\nabla \varphi (\vec r)$, the second derivative of a spherically
symmetric potential is:
\begin{equation}
 {d^2 \varphi (r) \over dr^2} = - {d g(r) \over dr} = G {d \over dr}
 \left ( m(r) { r \over (r^2)^{3/2} } \right ) \, ,
\end{equation}
where we now allow $r$ to take positive or negative values to
represent both sides of the origin. The density $\rho$ must remain
finite as $r \to 0$ and therefore, for a sufficiently small $r$ we may
approximate it by its value at the origin $\rho_0$ and then
$m(r)={4 \over 3}\pi r^3 \rho_0$. It is then straightforward to show that
\begin{equation}
  { d^2 \varphi \over dr^2}(r=0) = {4 \over 3} \pi \rho_0 \, .
\end{equation}
If $\rho$ is positive, then the potential $\varphi(r)$ is a concave
function (it has a minimum at $r=0$) and the central position is
stable against perturbations. For negative $\rho$ on the other hand,
we have the opposite behavior. The potential $\varphi(r)$ has a
maximum at $r=0$ and is a convex function of $r$. This latter
situation is what we have in the simulation and explains the
large-scale runaway effect.

\begin{figure}
  \resizebox{\hsize}{!}{\includegraphics{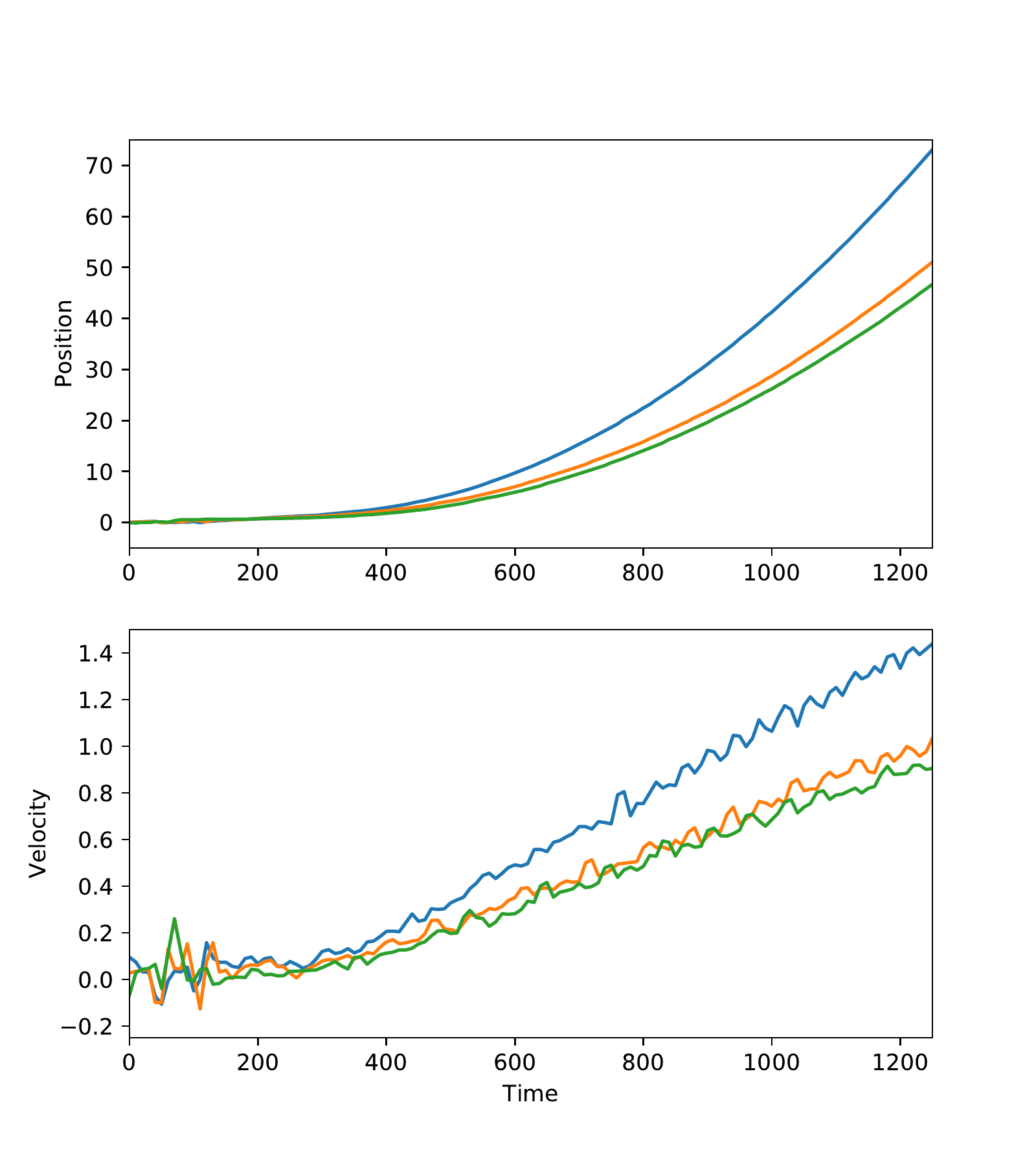}}
  \caption{ \label{fig:runaway} Upper panel: Evolution of the galactic
    core center-of-mass position as a function of time during the
    simulation run. The blue, orange, and green curves represent the
    position in $x$, $y,$ and $z$, respectively. Lower panel: As
    above but in this case referring to the velocity vector
    components. After an initial period of time at rest while the halo
    is building up, the entire system starts accelerating (at
    approximately $t=300$) and continues to gain speed with a constant
    acceleration. The sign of the $x$ and $y$ components has been
    reversed for better visualization in the figure.}
\end{figure}

Figure~\ref{fig:runaway} shows the temporal evolution of both position
and velocity of the galactic center of mass. Initially, the galaxy is
at rest while the halo density builds up. After some time, at
approximately $t=300$ the halo is sufficiently dense to push the
galaxy. The entire system starts accelerating and moves through the
box with a sustained acceleration, as evidenced by the constant slope
of the velocity curves in the lower panel.

\subsection{Halo shape}
\label{sec:shape}

The shape of the halo produced in the negative-mass simulations is
also problematic. Figure~\ref{fig:simhalo} (lower panel) already suggests
that the distribution of red dots around the galaxy is not
spherical. In particular, it is very elongated in the direction of the
runaway motion. This is clearly seen in the simulation video
available online\footnote{The video uses the same color scheme as F18}.

\begin{figure}
  \resizebox{\hsize}{!}{\includegraphics{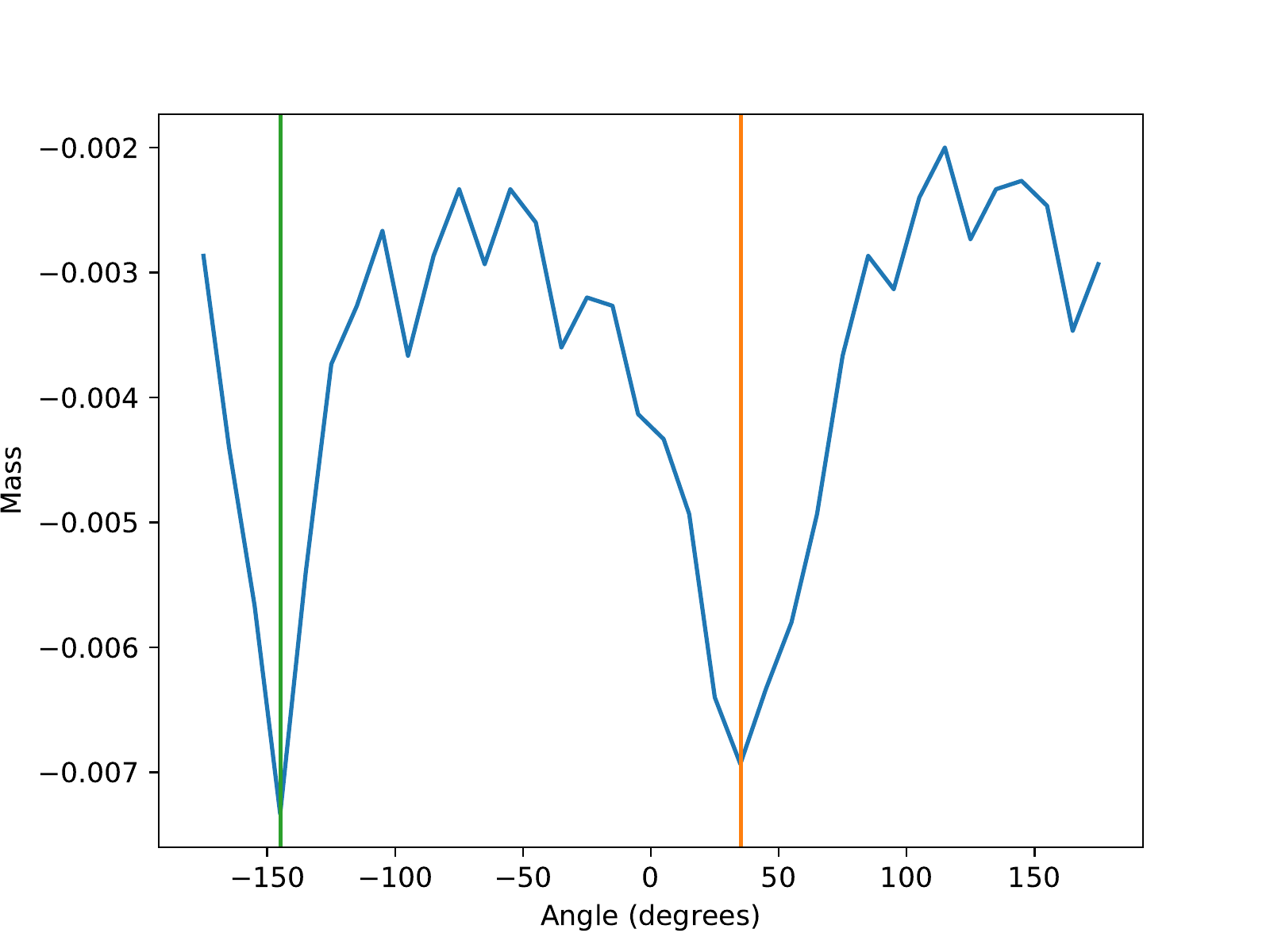}}
  \caption{ \label{fig:azim} Distribution of mass in the galactic halo
    integrated in bins of 10 degrees around the galactic
    center of mass in the $xy$-plane. The green (orange) line marks the
    direction ahead (behind) of the galactic runaway motion. }
\end{figure}

Figure~\ref{fig:azim} shows the distribution of halo mass
enclosed in bins of azimuth around the galactic center at the end of
the simulation. The vertical lines represent the instantaneous
direction of motion, with green and orange representing the directions
ahead and behind the galaxy. As seen in the figure, the halo is
extremely elongated along the galactic motion, slightly narrower ahead
and broader behind the galaxy.

\subsection{Cosmological structure formation}
\label{sec:cosmological}

\begin{figure}
  \resizebox{\hsize}{!}{\includegraphics{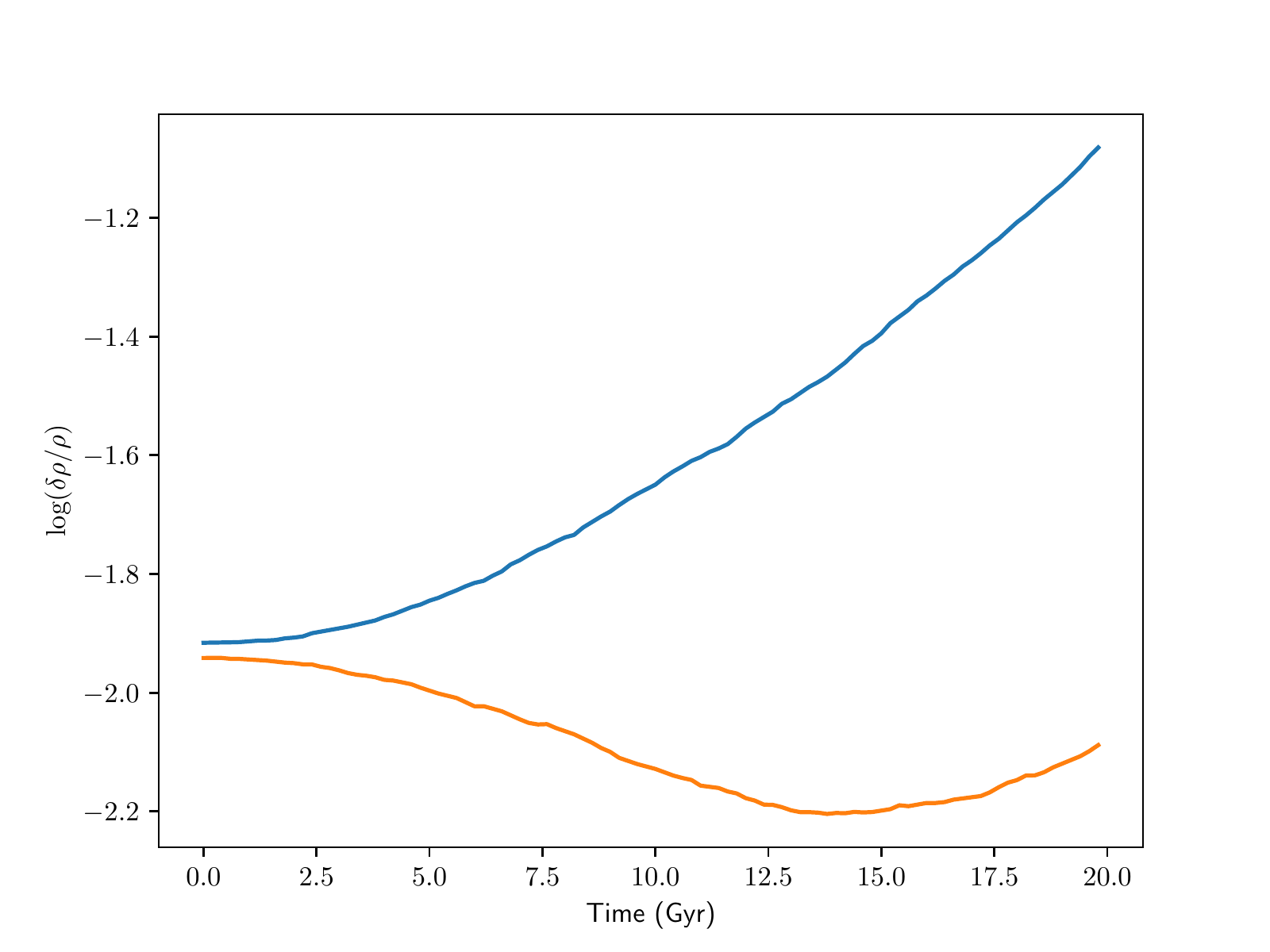}}
  \caption{ \label{fig:cosmological} Structure formation in a
    cosmological simulation. The plot represents the time evolution of
    relative overdensities, defined as the standard deviation of
    density variations among subdomains of 1~cubic Mlyr. Blue: Run
    with positive masses. Orange: Run with 84\% of negative masses.  }
\end{figure}

The choice of boundary conditions is most critical in the simulation
of structure formation inside a cosmological volume. Open boundary
conditions result in a much more rapid gravitational collapse. The
periodic boundary conditions imposed in the simulations presented here
are more realistic, with each point feeling approximately the same
gravitational pull from all directions (at least initially, with a
homogeneous density initialization), except for the starting small
random density fluctuations. Figure~\ref{fig:cosmological} shows the
result of a structure formation simulation in a large cosmological
volume. The box is one billion light-years (307~Mpc) in each direction
and the time step is 10~Myr. A total of 50,000 particles are tracked
in the simulation. They are initialized following a uniform
distribution in the box with a total matter density of 
  2.6$\times$10$^{-27}$~kg~m$^{-3}$. This is approximately 0.3 times the
  cosmological critical density, which has a value of
  8.6$\times$10$^{-27}$~kg~m$^{-3}$ (derived from cosmological
  parameters of \citealt{PAA+15}). Two different runs are shown in
the figure. In the first simulation, all particles have positive
mass. In the second one, 84\% of particles (the observed fraction of
dark matter) have a negative mass. The positive mass simulation
  has the initial random
inhomogeneities amplified by gravity to form concentrations and
voids at an accelerated rate. The negative-mass simulation begins with
the opposite behavior. The negative masses are dominant in number and
mass. Their mutual repulsion in the initial overdensities creates an
outward pressure which dilutes them and tends to smooth out
inhomogeneities. In fact, this is a trick that is often employed to
produce smooth initial conditions for regular cosmological simulations
(\citealt{W94}). Later on, the positive masses begin to coalesce and
form structures, dragging some negative halos around them. Structure
formation occurs at a much slower rate in the negative-mass
cosmological model.

\section{Other considerations}
\label{sec:others}

General relativity does not explicitly preclude the existence of
negative mass but the dominant energy condition (\citealt{HE73sect43})
and the positive energy theorem (\citealt{SY81}) are difficult to
reconcile with a negative mass (at least in the ADM sense,
\citealt{ADM60}). \cite{MP14} argued that loopholes may be found by
relaxing the theorem assumptions and suggested that the best candidate
is the requirement for an asymptotically Minkowskian space-time. These latter authors
proved that for an asymptotically de Sitter space-time, negative-mass
Schwarzschild solutions exist that everywhere satisfy the dominant
energy condition. However, this result does not apply to the F18
cosmology, which is equivalent to an anti-de Sitter space-time (a
negative cosmological constant). Therefore, while the existence of
negative-mass particles is not ruled out, it remains to be proven that
they may coexist with the energy condition in a background anti-de
Sitter space-time.

The runaway effect is certainly problematic. While it has been argued
that it does not violate momentum conservation, one needs to bear in
mind that momentum conservation was formulated assuming that mass is a
positive quantity. There are no physical, philosophical, or empirical
arguments to assume that it also holds when a mass is
negative. Extending the applicability of this principle to negative
masses is a generalization that should stand on its own merits. It
should not be considered as validated by an abuse of notation. A
particularly troublesome thought experiment was considered by Gold in
\cite{BBG+57}, in which one imagines attaching a pair of runaway
particles to a wheel, essentially building a perpetual motion machine
of the first kind.

If negative masses exist then it would be possible to build an
Alcubierre drive (\citealt{A94}). Such a device would enable
faster-than-light travel and, in consequence, closed timelike loops,
that is, the possibility to travel back in time (see,
e.g. \citealt{ER12}). A similar argument could be made about
  stabilizing traversable wormholes. Therefore, the existence of
negative masses would facilitate the occurrence of physical paradoxes
and causality violations.

The existence of a continuous negative-mass ocean of particles filling
the intergalactic space would impact the propagation of gravitational
waves from distant sources, which should lead to some effective
screening of the waves ( e.g., \citealt{MP14}). However, no
discrepancies have been found in the observation of several
gravitational wave events by the LIGO and VIRGO collaborations between
the wave signal amplitude and the theoretical expectation
(\citealt{AAA+18}). In particular, the GW170817 (kilonova)
  event, associated with an electromagnetic counterpart, yields the
  same distance when inferred from electromagnetic and gravitational
  waves (\citealt{AAA+17}).

\section{Conclusions}
\label{sec:conclusions}

The model proposed by F18 is a testable alternative theory of gravity
which if confirmed would have a profound impact on all areas of
physics. This paper presents a detailed analysis of his model, which
identifies a number of problems and discrepancies with
observations. Perhaps the most obvious weakness is the constantly
accelerating motion of galaxies. Instead of the orderly relation
between distance and velocity given by the Hubble-Lemaître law, we
would see all galaxies in the sky moving in random directions at
nearly the speed of light. Other serious problems include predicting
an incorrect shape, mass, and density profile for galactic halos, the
keplerian rotation curves, low mass of galactic halos, and slow
cosmological structure formation.

Other theoretical arguments
are discussed that are not directly testable with observations. For instance, the relationship between negative mass
and backwards time travel, opening the possibility of causality
violations, is an important issue for any theory or model involving
negative mass.

Overall, the conclusions of this study do not support the F18 model,
at least in the original formulation. Since the simulations are of
simplistic nature, as they were intended for exploratory purposes
following the philosophy of F18, it is possible that some of these
issues might be resolved with better, more realistic
simulations. Likewise, some of the theoretical objections, such as a
possible dampening of gravitational waves, would need actual
calculations to be formalized before they can be considered conclusive
evidence either for or against the theory.

\begin{acknowledgements}
The author gratefully acknowledges financial support from the Spanish
Ministry of Economy and Competitivity through project
AYA2014-60476-P. This research has made use of NASA's Astrophysics
Data System Bibliographic Services.  The Python Matplotlib\cite{H07},
Numpy\cite{numpy11} and iPython \cite{ipython07} modules have been
employed to generate the figures and calculations in this paper. 
  Thanks are also due to Isaac Alonso Asensio for his help identifying
  the F18 integration algorithm and implementing leapfrog in my code.
\end{acknowledgements}

\bibliographystyle{aa}
\bibliography{../bib/aamnem99,paper.bib,../bib/articulos}

\begin{thebibliography}{39}
\expandafter\ifx\csname natexlab\endcsname\relax\def\natexlab#1{#1}\fi

\bibitem[{{Abbott} {et~al.}(2017){Abbott}, {Abbott}, {Abbott}, {Acernese},
  {Ackley}, {Adams}, {Adams}, {Addesso}, {Adhikari}, {Adya}, \&
  et~al.}]{AAA+17}
{Abbott}, B.~P., {Abbott}, R., {Abbott}, T.~D., {et~al.} 2017, \apjl, 848, L12

\bibitem[{{Alcubierre}(1994)}]{A94}
{Alcubierre}, M. 1994, Classical and Quantum Gravity, 11, L73

\bibitem[{{Alemany} {et~al.}(2019){Alemany}, {Burrage}, {Bartosik}, {Bernhard},
  {Boyd}, {Brugger}, {Calviani}, {Carli}, {Charitonidis}, {Curtin}, {Dainese},
  {de Roeck}, {Diehl}, {D{\"o}brich}, {Evans}, {Feng}, {Ferro-Luzzi},
  {Gatignon}, {Gilardoni}, {Gninenko}, {Graziani}, {Gschwendtner}, {Goddard},
  {Hartin}, {Irastorza}, {Jaeckel}, {Jacobsson}, {Jungmann}, {Kirch}, {Kling},
  {Krasny}, {Lamont}, {Lanfranchi}, {Lansberg}, {Lindner}, {Long}, {Magnon},
  {Mallot}, {Martinez Vidal}, {Moulson}, {Papucci}, {Pawlowski}, {Pedraza},
  {Petridis}, {Pospelov}, {Pulawski}, {Redaelli}, {Rozanov}, {Rumolo}, {Ruoso},
  {Schacher}, {Schnell}, {Schuster}, {Semertzidis}, {Siemko}, {Spadaro},
  {Stapnes}, {Stocchi}, {Str{\"o}her}, {Usai}, {Vall{\'e}e}, {Venanzoni},
  {Wilkinson}, \& {Wing}}]{ABB+19}
{Alemany}, R., {Burrage}, C., {Bartosik}, H., {et~al.} 2019, arXiv e-prints,
  arXiv:1902.00260

\bibitem[{Arnowitt {et~al.}(1960)Arnowitt, Deser, \& Misner}]{ADM60}
Arnowitt, R., Deser, S., \& Misner, C.~W. 1960, Ann. Phys.(NY), 11, 116

\bibitem[{{Benoit-L{\'e}vy} \& {Chardin}(2012)}]{BLC12}
{Benoit-L{\'e}vy}, A. \& {Chardin}, G. 2012, \aap, 537, A78

\bibitem[{{Bondi}(1957)}]{B57}
{Bondi}, H. 1957, Reviews of Modern Physics, 29, 423

\bibitem[{Bondi {et~al.}(1957)Bondi, Bergmann, Gold, \& Pirani}]{BBG+57}
Bondi, H., Bergmann, P., Gold, T., \& Pirani, F. 1957, in The Role of
  Gravitation in Physics, Report from the 1957 Chapel Hill Conference, ed.
  D.~DeWitt, Cécile;~Rickles

\bibitem[{{Bondi} \& {Gold}(1948)}]{BG48}
{Bondi}, H. \& {Gold}, T. 1948, \mnras, 108, 252

\bibitem[{{Bonnor}(1989)}]{B89}
{Bonnor}, W.~B. 1989, General Relativity and Gravitation, 21, 1143

\bibitem[{Einstein(1918)}]{E18}
Einstein, A. 1918, The collected papers of {Albert Einstein}. {Vol}. 7, The
  {Berlin} years: writings, 1918--1921 (Princeton, NJ, USA: Princeton
  University Press), 33--36

\bibitem[{Everett \& Roman(2012)}]{ER12}
Everett, A. \& Roman, T. 2012, Time Travel and Warp Drives: A Scientific Guide
  to Shortcuts Through Time and Space (University of Chicago Press)

\bibitem[{{Farnes}(2018)}]{F18}
{Farnes}, J.~S. 2018, \aap, 620, A92

\bibitem[{Gauss(1813)}]{G1813}
Gauss, C.~F. 1813, Commentationes Societatis regiae scientiarum Gottingensis
  recentiores, Vol.~1 (Kraus Reprint), 355--378

\bibitem[{Hawking \& Ellis(1973)}]{HE73sect43}
Hawking, S.~W. \& Ellis, G. F.~R. 1973, {The Large Scale Structure of
  Space-Time}, Cambridge Monographs on Mathematical Physics (Cambridge
  University Press), 88

\bibitem[{{Hernquist}(1990)}]{H90}
{Hernquist}, L. 1990, \apj, 356, 359

\bibitem[{{Hossenfelder}(2008)}]{H08}
{Hossenfelder}, S. 2008, \prd, 78, 044015

\bibitem[{{Hubble}(1929)}]{H29}
{Hubble}, E. 1929, Proceedings of the National Academy of Science, 15, 168

\bibitem[{Hunter(2007)}]{H07}
Hunter, J.~D. 2007, Computing In Science \& Engineering, 9, 90

\bibitem[{{Lema{\^i}tre}(1927)}]{L27}
{Lema{\^i}tre}, G. 1927, Annales de la Soci{\'e}t{\'e} Scientifique de
  Bruxelles, 47, 49

\bibitem[{Luttinger(1951)}]{L51}
Luttinger, J.~M. 1951, On "Negative" mass in the theory of gravitation (Awards
  for Essays on Gravitation. Gravity Research Foundation)

\bibitem[{{Manfredi} {et~al.}(2018){Manfredi}, {Rouet}, {Miller}, \&
  {Chardin}}]{MGR+18}
{Manfredi}, G., {Rouet}, J.-L., {Miller}, B., \& {Chardin}, G. 2018, \prd, 98,
  023514

\bibitem[{{Mbarek} \& {Paranjape}(2014)}]{MP14}
{Mbarek}, S. \& {Paranjape}, M.~B. 2014, \prd, 90, 101502

\bibitem[{{MPI Forum}(1994)}]{mpi94}
{MPI Forum}. 1994, MPI: A Message-Passing Interface Standard, Tech. rep.,
  Knoxville, TN, USA

\bibitem[{{Navarro} {et~al.}(1996){Navarro}, {Frenk}, \& {White}}]{NFW96}
{Navarro}, J.~F., {Frenk}, C.~S., \& {White}, S.~D.~M. 1996, \apj, 462, 563

\bibitem[{Newton(1833)}]{N1833}
Newton, I. 1833, Philosophiae naturalis principia mathematica, Vol.~1 (G.
  Brookman)

\bibitem[{{Peebles} \& {Ratra}(2003)}]{PR03}
{Peebles}, P.~J. \& {Ratra}, B. 2003, Reviews of Modern Physics, 75, 559

\bibitem[{{Penrose}(1969)}]{P69}
{Penrose}, R. 1969, Nuovo Cimento Rivista Serie, 1

\bibitem[{{Penzias} \& {Wilson}(1965)}]{PW65}
{Penzias}, A.~A. \& {Wilson}, R.~W. 1965, \apj, 142, 419

\bibitem[{P\'erez \& Granger(2007)}]{ipython07}
P\'erez, F. \& Granger, B.~E. 2007, Computing in Science and Engineering, 9, 21

\bibitem[{{Petit} \& {d'Agostini}(2014)}]{PdA14}
{Petit}, J.~P. \& {d'Agostini}, G. 2014, \apss, 354, 611

\bibitem[{{Planck Collaboration} {et~al.}(2016){Planck Collaboration}, {Ade},
  {Aghanim}, {Arnaud}, {Ashdown}, {Aumont}, {Baccigalupi}, {Banday},
  {Barreiro}, {Bartlett}, \& et~al.}]{PAA+15}
{Planck Collaboration}, {Ade}, P.~A.~R., {Aghanim}, N., {et~al.} 2016, \aap,
  594, A13

\bibitem[{{Riess} {et~al.}(1998){Riess}, {Filippenko}, {Challis},
  {Clocchiatti}, {Diercks}, {Garnavich}, {Gilliland}, {Hogan}, {Jha},
  {Kirshner}, {Leibundgut}, {Phillips}, {Reiss}, {Schmidt}, {Schommer},
  {Smith}, {Spyromilio}, {Stubbs}, {Suntzeff}, \& {Tonry}}]{RFC+98}
{Riess}, A.~G., {Filippenko}, A.~V., {Challis}, P., {et~al.} 1998, \aj, 116,
  1009

\bibitem[{{Rubin} \& {Ford}(1970)}]{R70}
{Rubin}, V.~C. \& {Ford}, Jr., W.~K. 1970, \apj, 159, 379

\bibitem[{Schoen \& Yau(1981)}]{SY81}
Schoen, R. \& Yau, S.~T. 1981, Comm. Math. Phys., 79, 231

\bibitem[{{Schwarzschild}(1916)}]{S16}
{Schwarzschild}, K. 1916, Abh.~Konigl.~Preuss.~Akad.~Wissenschaften Jahre
  1906,92, Berlin,1907, 1916

\bibitem[{{Takahashi} \& {Asada}(2013)}]{TA13}
{Takahashi}, R. \& {Asada}, H. 2013, \apjl, 768, L16

\bibitem[{{The LIGO Scientific Collaboration} \& {the Virgo
  Collaboration}(2018)}]{AAA+18}
{The LIGO Scientific Collaboration} \& {the Virgo Collaboration}. 2018, arXiv
  e-prints, arXiv:1811.12907

\bibitem[{Van Der~Walt {et~al.}(2011)Van Der~Walt, Colbert, \&
  Varoquaux}]{numpy11}
Van Der~Walt, S., Colbert, S.~C., \& Varoquaux, G. 2011, Computing in Science
  \& Engineering, 13, 22

\bibitem[{{White}(1994)}]{W94}
{White}, S. D.~M. 1994, arXiv e-prints, astro

\end{thebibliography}

\end{document}